\documentclass[fleqn,usenatbib]{mnras}

\usepackage{newtxtext,newtxmath}

\usepackage[T1]{fontenc}

\DeclareRobustCommand{\VAN}[3]{#2}
\let\VANthebibliography\thebibliography
\def\thebibliography{\DeclareRobustCommand{\VAN}[3]{##3}\VANthebibliography}

\usepackage{graphicx}	
\usepackage{amsmath}	

\title[New satellite galaxy candidates of NGC2683]{New dwarf galaxy candidates in the sphere of influence of the Local Volume spiral galaxy NGC2683}

\author[E.~Crosby et al.]{Ethan Crosby$^{1}$\thanks{E-mail: Ethan.Crosby@anu.edu.au},
Helmut Jerjen$^{1}$,
Oliver M\"uller$^{2}$,
Marcel Pawlowski$^{3}$, Mario Mateo$^{4}$\newauthor
and Markus Dirnberger$^{1}$
\\
$^{1}$Research School of Astronomy and Astrophysics, Australian National University, Canberra, ACT 2611, Australia\\
$^{2}$Institute of Physics, Laboratory of Astrophysics, Ecole Polytechnique F\'ed'erale de Lausanne (EPFL), 1290 Sauverny, Switzerland\\
$^{3}$Leibniz-Institut f\"ur Astrophysik Potsdam, An der Sternwarte 16, D-14482 Potsdam, Germany\\
$^{4}$ Department of Astronomy, University of Michigan, 1085 S. University Ave., Ann Arbor, MI 48109, USA \\
}
\date{Accepted XXX. Received YYY; in original form ZZZ}

\pubyear{2022}

\begin{document}
\label{firstpage}
\pagerange{\pageref{firstpage}--\pageref{lastpage}}
\maketitle

\begin{abstract}

We present initial results of a survey of host $L_{*}$ galaxies environments in the Local Volume ($D<10\,$Mpc) searching for satellite dwarf galaxy candidates using the wide-field Hyper Suprime-Cam imager on the 8m Subaru Telescope. The current paper presents complete results on NGC2683 ($M_{B_T,0}=-19.62$, $D=9.36\,Mpc$, $v_{\odot}=411\,km\,s^{-1}$), an isolated Sc spiral galaxy in the Leo Spur. At the distance of NGC2683, we image the complete volume out to projected radii of $380\,kpc$ using a hexagonal arrangement of 7 pointings. Direct inspection of the images is complete down to $M_{g}\sim-11$ and has revealed 4 new satellite galaxy candidates, 2 of which have been independently discovered by other researchers. Assuming the distance of NGC2683, these candidates span luminosities $-12 < M_g < -9$ and effective radii 150\,pc $< r_e <$ 1100\,pc and are found to be morphologically reminiscent of satellite galaxies in the Local Group. These 4 new candidates add to the 8 already known. A Principle Component Analysis of the 2D projected distribution of the 12 satellite galaxies of NGC2683 reveals a flattened projected disk of satellites, with axis ratio $b/a=0.23$. This flattening in the 2D projected system of satellites is a 1 percent outlier of simulated isotropic satellite systems but is mostly consistent with satellite distributions of comparable galaxy environments in the IllustrisTNG simulation. This indicates the possible presence of a satellite plane, which will need to be investigated with follow-up observations.

\end{abstract}



\section{Introduction}

\subsection{Background}
The Lambda Cold Dark Matter ($\Lambda$CDM) model is the predominant and generally accepted cosmological model predicting the formation of galactic structures and has been well-studied through observations \citep{Planck2016,Abbott_et_al_2018,Planck2020} and simulations \citep{Springel2008, Vogelsberger2014,EAGLE_2015,TNG100,Alam2021,Panth2021}. In the $\Lambda$CDM paradigm dark matter forms gravitationally bound halos, the larger of which are often accompanied by numerous bound smaller dark matter subhalos. These halos generally accrete sufficient baryonic matter to form large galaxies and companion satellite dwarf galaxies respectively \citep{Moore1999,Klypin1999,Springel2008, GarrisonKimmel2014,Griffen2016,Kelley2019}. The comparison between high-resolution cosmological simulations and the observed universe provides critical insight into the relative importance of the various physical mechanisms that trigger or suppress star-formation in the models of the nearby universe.

Such comparisons have historically revealed a number of interesting and surprising discrepancies between observations and simulations. Here we focus on issues at galaxy scales ($<1$\,Mpc), particularly relating to satellite galaxy systems of single $L^*$ host galaxy and small galaxy group environments. These issues include the ``Missing Satellites", ``Core-Cusp", and ``Too-Big-To-Fail" (\textit{TBTF}) problems (see \cite{Bullock2017} for a review), but have since at least been alleviated by the addition of resolved baryon physics in cosmological zoom-in simulations \citep{Buck2018,GarrisonKimmel2019,Wheeler2019}. Ongoing tensions for these problems still exist, but it is indicated at least part of the problem lies in resolved baryon simulations.

The inclusion of baryon physics however does not resolve a fourth problem known as the ``Disk-of-Satellites" or ``Satellite Plane" phenomenon. A satellite plane is the observation of a co-moving, aligned flattened distribution of satellite galaxies which is known to exist in the Local Group \citep{LyndenBell1976,Kroupa2005,Metz2007,Pawlowski2012,Ibata2013,Conn2013} and the Centaurus A/M83 Group \citep{Tully2015,Mueller2018,Mueller_2019b,Muller_2022}. Indications of satellite planes also exist in M81 \citep{Chiboucas2013}, M101 \citep{Mller2017} and NGC 253 \citep{MartnezDelgado2021}. All of these host galaxies together reside in the Local Sheet, a large-scale planar structure of nearby galaxy groups which are collectively moving towards the Virgo cluster \citep{Tully2008,Libeskind_2015}. 

The observations of anisotropic, flattened satellite systems clash with results from $\Lambda$CDM simulations which do not commonly produce planar alignments of satellite galaxies about hosts, let alone display co-orbiting behaviour \citep{Pawlowski_2021}. A number of solutions have been put forward, firstly of a physical nature in which satellite galaxies are formed or accreted within the $\Lambda$CDM paradigm. Such proposals include galaxy accretion along flattened or narrow cosmic filaments \citep{Libeskind2010,Lovell2011}, group infall of satellites \citep{Metz2009} (see \citealt{Pawlowski2018} for a review). A second possible reason for the extraordinary observations of planar satellite systems is the \textit{"look-elsewhere"} effect. In this case, the purely statistical phenomenon of selection bias formed from limited sample sets which are not truly independent leads to misleading, inflated statistical significance \citep{Cautun2015}. 

In both cases, it is implied that cosmological structures such as the Local Sheet within which many of the previously studied galaxy systems lie, may have either provided the physical conditions necessary for forming these satellite planes or created a biased statistical sample, perhaps even both. However it is notable that even in this scenario of filamentary accretion, well-studied systems, particularly the Local Group still lie in the statistical tail of the distribution of satellite system alignments: only $0.1\sim10$ percent of $\Lambda$CDM hydro-dynamical simulated systems produce such observed alignments \citep{Pawlowski2012,Pawlowski2013,Ibata_2014,ForeroRomero2018}.

The observation of unexpected large scale planar satellite structures begs the question: are the formation conditions of galaxies in the Local Sheet truly a significant outlier, or do $\Lambda$CDM simulations fail to account for an unknown factor relating to the formation of such planar structures? Answering this question requires a robust sample of satellite galaxy systems that minimises selection bias.

\subsection{This Work}
In this paper we present the detections and fundamental properties of satellite galaxy candidates around NGC2683 using deep Subaru Hyper Suprime-Cam imaging data. NGC2683 (PGC24930) is a luminous, highly inclined Sc spiral ($M_{B_T,0}=-20.24$) at a distance of $9.36\pm0.28$\,Mpc located in the Leo Spur, a sparsely populated, filamentary structure beyond the Local Sheet, but within the Local Volume, a spherical region of space with a radius of $\approx 10$\,Mpc around the Milky Way \citep{Karachentsev_2015_A}. Galaxies in the Leo Spur are known to have relatively large peculiar velocities towards the Milky Way. NGC2683 has a measured heliocentric velocity of v$_\odot=411$\,km\,s$^{-1}$ \citep{Haynes_1998} and thus a peculiar velocity of the order of $-220$\,km\,s$^{-1}$ assuming $H_0=67.4$\,km\,s$^{-1}$\,Mpc$^{-1}$ \citep{Planck2020}. The virial radius of NGC2683 was estimated using the relation:
\begin{equation}
	R_{200} = \sqrt[3]{\frac{3M_{200}}{4\pi\left(200\rho_{crit}(z)\right)}}
\end{equation}
from \citet{Kravtsov_2013} where $R_{200}$ is assumed to be the virial radius, $M_{200}$ the mass within that radius and $200\rho_{crit(z)}$ is 200 times the critical density of the universe as a function of redshift. $M_{200}$ is estimated using the stellar mass ($M_{*}$) in Table \ref{tab:NGC2683param} \citep[from][]{Vollmer_2016} and the average $M_{*}/M_{200}$ ratio from select galaxies in the \emph{TNG100-1} model, where the selection criteria and model is described in more detail in Section \ref{NGC2683_sat_plane}. We calculated the critical density at $z=0$ using the expression:
\begin{equation}
	\rho_{crit}(z=0) = \frac{3H_0^2}{8\pi\,G}
\end{equation}
Where we assume the Hubble constant $H_0$ is that measured by \citet{Planck2020} ($H_0=67.4$\,km\,s$^{-1}$\,Mpc$^{-1}$), in order to be consistent with large scale cosmological simulations. We estimate the virial radius of NGC2683 to be $R_{200}\approx 220$\,kpc.
A complete list of fundamental parameters for NGC2683 is given in Table\,\ref{tab:NGC2683param}. 

\begin{table}
      \caption{Fundamental parameters of the host galaxy NGC~2683}
         \label{tab:NGC2683param}
         \begin{tabular}{lll}
           \hline
           Morphology & Sc & \cite{Ann_2015} \\
           R.A.(J2000) & 08:52:41.3& \\
           DEC (J2000) & +33:25:18 &          \\
           Inclination & $42^\circ.6$  &          \\
           v$_\odot$ & 411\,km\,s$^{-1}$ & \cite{Haynes_1998} \\
           m$_{\rm B}^{0}$ & $10.24$~mag & \cite{RC3}\\
           $D_{26}$ & $9\farcm 3=26.1\,$kpc & \cite{RC3} \\
           Distance & $9.36\pm0.28\,$Mpc & \cite{Karachentsev_2015_A}\\
        $(m-M)$ & $29.86\pm0.06$ & \cite{Karachentsev_2015_A}\\
           $M_{B_T,0}$ & $-19.62$~mag & this study\\
           $v_{\rm rot}^{\rm max}$ & $205$~km\,s$^{-1}$ & \cite{Casertano1991} \\ 
           $M_{*}$ & $3.6 \times 10^{10}$~M$_{\odot}$  & \cite{Vollmer_2016}\\
           $M_{200}$ & $1.2 \times 10^{12}$~M$_{\odot}$ & this study\\
           $R_{200}$ & $220\,$kpc\ & this study \\
           \hline
         \end{tabular}
\end{table}

\section{Observations}
We obtained CCD images of the NGC2683 region in the \textit{HSC-g} band using the Hyper Suprime Camera \citep[HSC,][]{2018PASJ...70S...1M} at the 8.2m Subaru telescope at the Mauna Kea Observatories. The data acquisition was conducted as part of the observing proposal S18B0118QN (PI: H. Jerjen) on 2019 January 30-31. This proposal called for accompanying \textit{HSC-r2} imaging, however this could not be completed due to poor weather conditions during one night. The average seeing for the \textit{HSC-g} band observations was $1\farcs26\pm 0\farcs39$ arcsec. The HyperSuprime Camera is equipped with an array of 104 4k$\times$2k science CCD detectors, with an angular diameter of 1.5 degrees and a pixel scale of $0\farcs169$ at the centre of the field \citep{2018PASJ...70S...1M}. At the distance of NGC2683 this angular diameter corresponds to a physical size of $\sim253\,$kpc.

\begin{figure*}
	\includegraphics[draft=false,width=16cm]{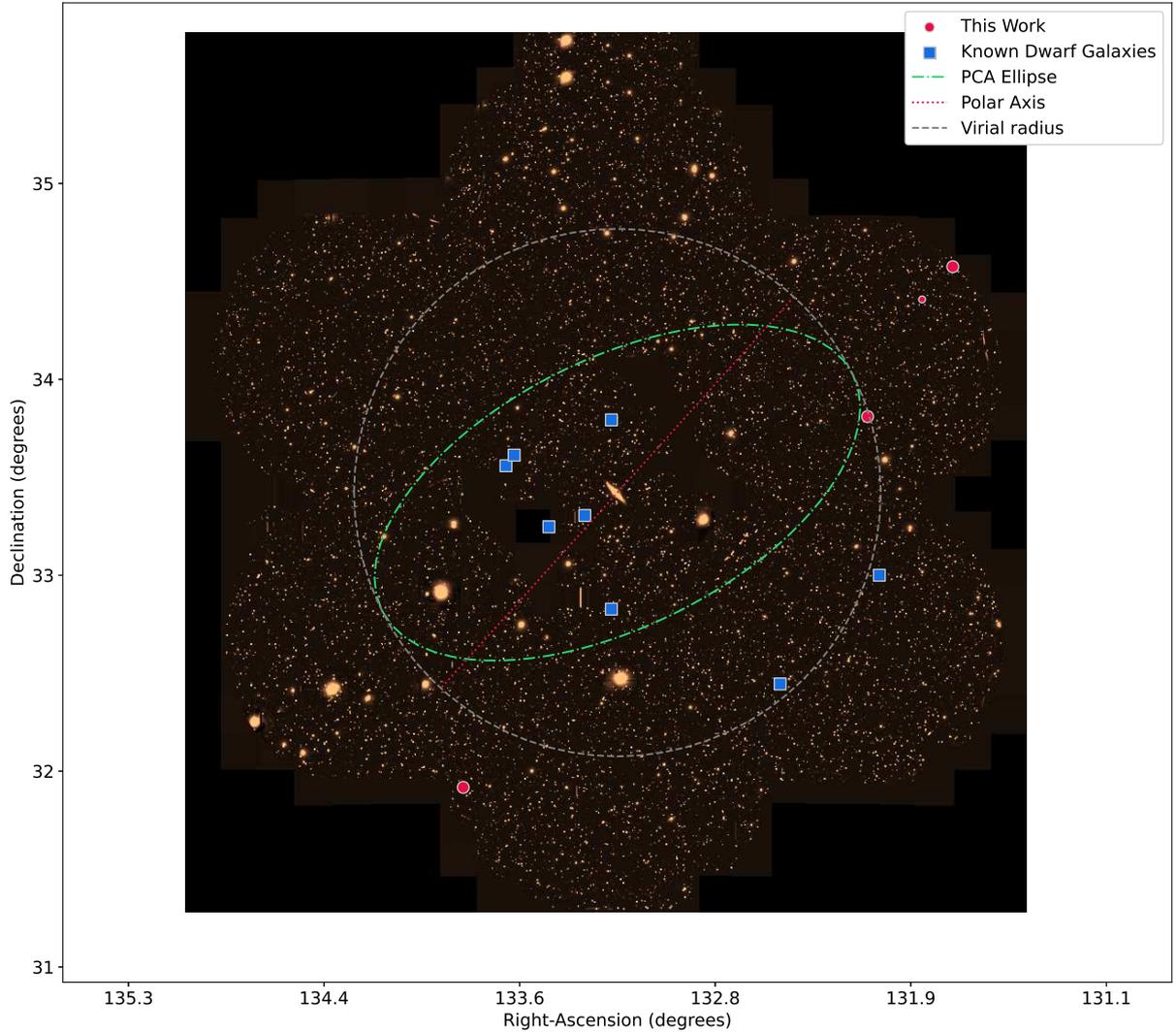}
	\caption{Map showing the HSC survey footprint centred around the edge-on Sc spiral galaxy NGC2683. The locations of previously known NGC2683 dwarf satellite galaxy candidates are shown as blue squares, while new candidates from this work are the red circles. The size of the symbols is proportional to the NGC2683 group membership probability as determined in our analysis. Dark black areas are outside of the region captured by the telescope. The grey dashed line represents the approximate virial radius of NGC2683, the green dashed-dotted line the best fit PCA ellipse of the 2D satellite distribution as described in Section \ref{NGC2683_sat_plane} and the red dotted line the polar axis of NGC2683. The major axis of the PCA ellipse and the polar axis are separated by $28$ degrees.}
	\label{fig:NGC2683_coverage}
\end{figure*}

During the observing run, seven HSC pointings were observed in a hexagonal pattern, with the central field on NGC2683, extending the survey area out to a complete spherical volume with radius $\sim380\,$kpc at the distance of NGC2683, far beyond the estimated virial radius of NGC2683 of $220\,$kpc. Figure \ref{fig:NGC2683_coverage} shows the survey footprint, as well as known dwarf galaxies and new satellite candidates of NGC2683. This observing configuration was chosen to achieve our scientific goal of detecting low surface brightness dwarf galaxies out to the virial radius of NGC2683, establishing a dataset that, with follow-up observations, can be used to test planar alignment and co-rotation of satellite galaxies in this system. Each HSC pointing consisted of three exposures: a short 30\,sec exposure for photometric calibration purposes and two 150\,sec exposures, dithered by half a CCD in R.A. and DEC directions.
This strategy allows us to detect satellite galaxies and low surface brightness dwarf galaxies down to a mean effective surface brightness of ${\sim\,26.5\,\text{mag}\,\text{arcsec}^{-2}}$ as shown in section \ref{sec_completeness}. 

\section{Data reduction}
The data was processed using the open-source  {\it hscpipe} software, which is based on the pipeline being developed for the LSST Data Management system. We work with fully reduced, sky-subtracted co-add images, which  {\it hscpipe} divides into equal rectangular regions called \textit{patches} of $4200\times4200$ pixels. These patches contain a 100 pixel overlap of neighbouring patches on all sides, to minimise any selection bias introduced by the arbitrary choice of patch boundaries. 

Astrometric and photometric calibrations were performed relative to the Pan-STARRS1 reference catalogue \citep{ps1_eprint}. An in-depth discussion of the established data reduction procedure for the {\it hscpipe} is presented by \cite{Bosch2018}. For our dataset, we follow a similar quality assurance process to the Hyper Suprime-Cam Subaru Strategic Program \citep{Aihara2019}. For the purposes of constructing a photometric error budget, we interpret the uncertainty in the photometric zeropoint as $\sigma_{\text{ext}} = 0.087$\,mag.

Given that we are looking for low-surface brightness galaxies, which are by definition fainter than the night sky, the high quality sky subtraction is crucial for our search. We are using a new and improved sky-subtraction algorithm included with {\it hscpipe} 7.9.1, which performs a global sky subtraction based on an empirical background model extending over the entire focal plane. As this model operates across the boundaries of individual CCDs, this eliminates effects arising from the discontinuities of CCD edges. Furthermore, a scaled sky frame - the mean response of the instrument to the sky for a particular filter - is used. This allows static features to be subtracted at smaller scales than the empirical model, and further improves the photometry of our observations.

\section{Known satellite galaxies}
Previous studies have been conducted to search for satellite galaxies around NGC2683. Before our survey, eight candidates have been identified and imaged based on optical observations \citep{Karachentseva_1998,Karachentsev_2015_A, Javanmardi_2016,Carlsten_2022} and an additional two candidates we present here were independently discovered in \citealt{Carlsten_2022}. Only two satellites have group membership confirmed with a distance, via tip magnitude of the red giant branch (TRGB). An additional object, a HI cloud close to NGC2683 without optical counterpart was detected at radio frequencies \citep{Saponara_2020}. Photometric, structural and spatial parameters are given in Table\,\ref{tab:photometry} and images of these satellites as presented in Fig.\ref{fig:known_satellites_pics}. Four of the eight objects presented here are not covered by our HSC dataset due to problems with the data processing in the region of the image containing those galaxies. These objects are KK70, NGC2683 dw1, NGC2683 dw2 and NGC2683 DGSAT-3. In these regions, we instead verified existing candidates using \textsc{HSCLA2016} data from the HSC Legacy Archive \citep{HSC_leg} which process images that partially cover our footprint as shown in Fig.\ref{fig:NGC2683_coverage}. This data is only in the \textit{HSC-r} band and is shallower, ie. has a higher surface brightness limit by about 0.5-1.0 magnitudes, than our stacked and processed \textit{g} band data.

\subsection{KK[98a]69, NGC2683-DGSAT-1 or dw0857p3347}
This irregular, low-surface brightness galaxy (top left panel in Fig.\ref{fig:known_satellites_pics}) is the most luminous of the known NGC2683 satellites \citep{Karachentseva_1998}. It has an \textit{r} band luminosity of $M_{r}=-15.56$ \citep{Javanmardi_2016},
a heliocentric velocity of 463\,km\,s$^{-1}$ \citep{Saponara_2020} and a tip of the red giant branch distance of $9.28\pm0.28$\,Mpc \citep{Karachentsev_2015_A}. Follow-up observation at 21cm with the Giant Metrewave Radio Telescope as part of the FIGGS survey \citep{Begum2008} has also revealed an asymmetric and off-centred HI distribution with an estimated total neutral hydrogen mass of $M_{HI}=4.2\times 10^{7}M_\odot$ \citep{Saponara_2020}. This suggests the galaxy may be properly classified as a transitional dwarf galaxy given the sluggish star formation rate of $1.6\times10^{-4}\,M_{\odot}yr^{-1}$ \citep{Karachentsev_2010} and the presence of a displaced HI cloud \citep{Saponara_2020}, properties that resemble the Phoenix dwarf which is likewise classified as a transitional dwarf galaxy \citep{Young2007}. Additionally, KK69 has a very low central surface brightness $\mu_{g,0}=25.4\,\text{arcsec}^{-2}$ and extended half-light radius $r_{e}=2.07\,kpc$, which meets the general classification criteria of an Ultra Diffuse Dwarf (UDG) \citep{van_Dokkum_2015} and may indicate that it is transitioning into such as object. KK[98a]69 is therefore is an example of a rare transitioning UDG that possesses an offset HI distribution. KK[98a]69 is the only one of the known NGC2683 satellites fully imaged by our HSC dataset and is invisible in the \textsc{HSCLA2016} dataset. Using \textsc{galfit} modelling as discussed in Section 6.3, we present photometric parameters for this galaxy based on our analysis in Appendix A.

\subsection{KK[98a]70, NGC2683-DGSAT-2 or dw0855p3333}
This dwarf galaxy, KK[98a]70, (top right panel in Fig.\ref{fig:known_satellites_pics}) is more compact with a 0.67\,mag higher effective surface brightness that KK[98a]69. KK[98a]70 has a total luminosity of $M_{r}=-13.94$ ($M_B=-12.3$).
The line-of-sight velocity remains unknown, but a TRGB distance of $9.18\pm0.30$\,Mpc has been measured by \cite{Karachentsev_2015_A}, confirming its group membership. The morphology resembles that of a dwarf spheroidal galaxy.

\subsection{NGC2683-DGSAT-3 or dw0855p3336}

An elliptical, dwarf galaxy (third row, left panel in Fig.\ref{fig:known_satellites_pics}) with an \textit{r} band luminosity of $M_{r}=-11.45$ \citep{Javanmardi_2016}. Without distance or velocity measurements its group membership still remains undetermined. We consider this object a high-probability candidate based on upon its smooth elliptical morphology strongly resembling an early type dwarf galaxy.

\subsection{NGC2683 dw2, NGC2683-DGSAT-4 or dw0854p3314}

A small dwarf spheroidal galaxy (second row, right panel in Fig.\ref{fig:known_satellites_pics}), with a half-light radius of 9.9 arcsec and an \textit{r} band luminosity of $M_{r}=-11.06$ \citep{Javanmardi_2016}. Without distance or velocity measurements the group membership still remains undetermined. We consider this object a high-probability candidate given its smooth morphology, resembling an early type dwarf galaxy.

\subsection{NGC2683-DGSAT-5 or dw0852p3249}

This is the smallest ($r_e=6.8$\,arcsec) spheroidal galaxy (third row, right panel in Fig.\ref{fig:known_satellites_pics}) of the known candidates, with an \textit{r} band luminosity of $M_{r}=-9.97$ \citep{Javanmardi_2016}. Without distance or velocity measurements the group membership still remains undetermined. We consider this object a high-probability candidate given its smooth morphology, resembling an early type dwarf galaxy.

\subsection{NGC2683 dw1, LV0853+3318, GALEXASC J085326.78+331818.3 or dw0853p3318}

This irregular galaxy (second row, left panel in Fig.\ref{fig:known_satellites_pics}) has H$\alpha$ and FUV GALEX emission data and has previously assumed to be a part of the NGC2683 system. With a recently measured recessional velocity of $421$\,km\,s$^{-1}$ \citep{Saponara_2020} determined from its HI emission, it is confirmed to be a group member. The estimated star-formation rate is $6.3\times10^{-4}\,M_{\odot}yr^{-1}$ \citep{Karachentsev_2015_B}, highest among all known satellites around NGC2683.

\subsection{dw0846p3300}

This is a disturbed, low surface brightness irregular galaxy (bottom left panel in Fig. \ref{fig:known_satellites_pics}) with a \textit{g} band luminosity of $M_{g}=-12.2$ and a half-light radius of $\sim2.6\,kpc$ from our \textsc{GALFIT} model (see Tab. \ref{tab:photometry}). This gives this object a central surface brightness of $\mu_{g,0}\,\sim\,26.8$ and a mean surface brightness within the effective radius of $\langle\mu_{g,e}\rangle\,\sim\,28.4$. Additionally, this object is not well-fit by a single S\'ersic profile and is thus best classified as a low-surface brightness irregular or ultra-diffuse galaxy. Given it's unusual morphology that is nearly unmistakably reminiscent of a dwarf galaxy, we assume this is a high probability satellite candidate.

\subsection{dw0848p3226}

\begin{figure}
	\centering
	\includegraphics[draft=false,width=8cm]{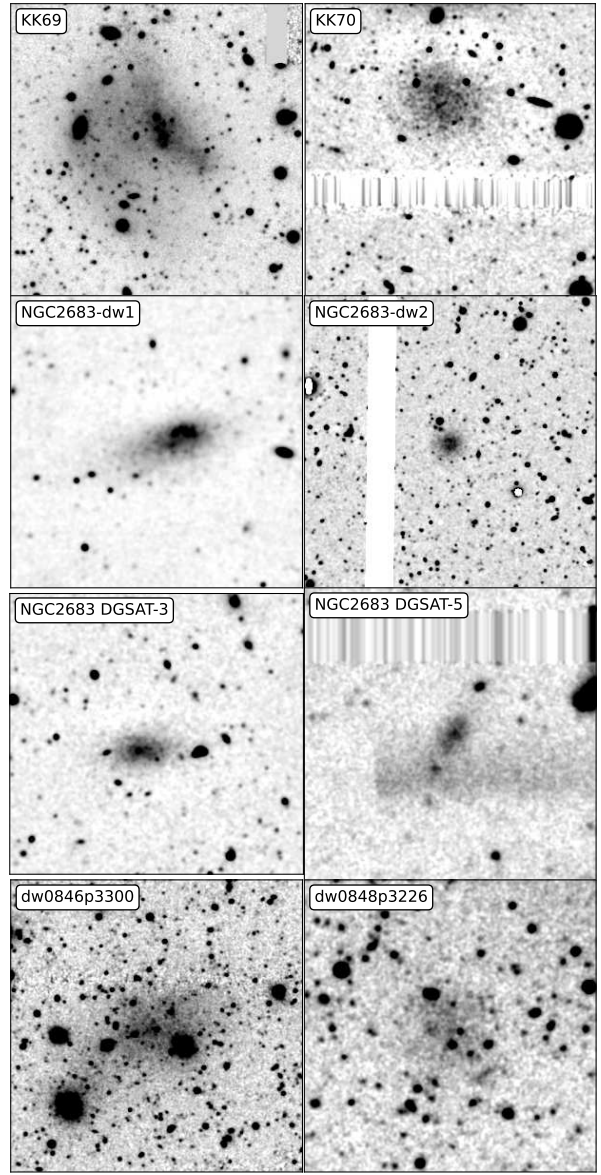}
	\caption{Images of all previously known NGC2983 satellite galaxies and candidates. KK69, NGC2683 DGSAT-5, dw0846p3300 and dw0848p3226 are produced from our \textit{g}-band data, the others from \textsc{HSCLA2016} data \citep{HSC_leg}. The images show the range of detected galaxy morphologies and surface brightness. From top left to bottom right: KK[98a]69, KK[98a]70, NGC2683 dw1, NGC2683 dw2, NGC2683-DGSAT-3, NGC2683-DGSAT-5, dw0846p3300 and dw0848p3226. No optical image exists for NGC2683 dw3? (see Section \ref{dw3_cloud}), which is most likely a HI cloud of the host galaxy NGC2683.}
	\label{fig:known_satellites_pics}
\end{figure}

This is low surface brightness dwarf spheroidal (bottom right panel in Fig.\ref{fig:known_satellites_pics}) with \textit{g} band luminosity of $M_{g}=-9.5$ from our \textsc{GALFIT} model (see Tab. \ref{tab:photometry}). Without known distance or velocity information, we assume this object is a high probability satellite candidate based on its early type dwarf galaxy morphology.

\subsection{NGC2683 dw3?} \label{dw3_cloud}

Although labelled a dwarf (dw), it is detected only at 21\,cm as an HI cloud with gas mass of $\approx 10^7 M_\odot$ and a radial velocity of 467\,km\,s$^{-1}$ \citep{Saponara_2020}. Though it is in close proximity to NGC2683 itself ($10\farcm 31$ or $28.1\,$kpc north-east of the NGC 2683 HI disk), \cite{Saponara_2020} hypothesised it is a satellite galaxy or possibly associated with KK69 due to the similar velocities of the two objects, despite being separated by $\sim45\,$kpc. It is approximately $2.4 \times 1.6\,$kpc in size. No optical counterpart of this object exists in data sets of other surveys or our own, which leads us to suggest it is not a satellite galaxy.

\section{Search for new satellite galaxies}

In order to find new satellite galaxy candidates in the NGC2683 Group, we search for unresolved, low surface brightness objects in our HSC observations. A fail safe way to detect these candidates is through visual inspection. We employ this approach through a meticulous independent inspection of the entire survey area by two authors. The entire data set is reviewed multiple times and possible objects are logged and categorised based on their morphology. The images of known dwarf galaxies in the NGC2683 system (Fig. \ref{fig:known_satellites_pics}) serve as a guidance to the appearance of potential new satellite galaxies. Our strategy is a conservative one, we only present satellite galaxies here that are unlikely to be false-positive candidates.

\subsection{Visual detection of dwarf galaxy candidates} \label{visual_det}

\begin{figure*}
	\centering
	\includegraphics[draft=false,width=18cm]{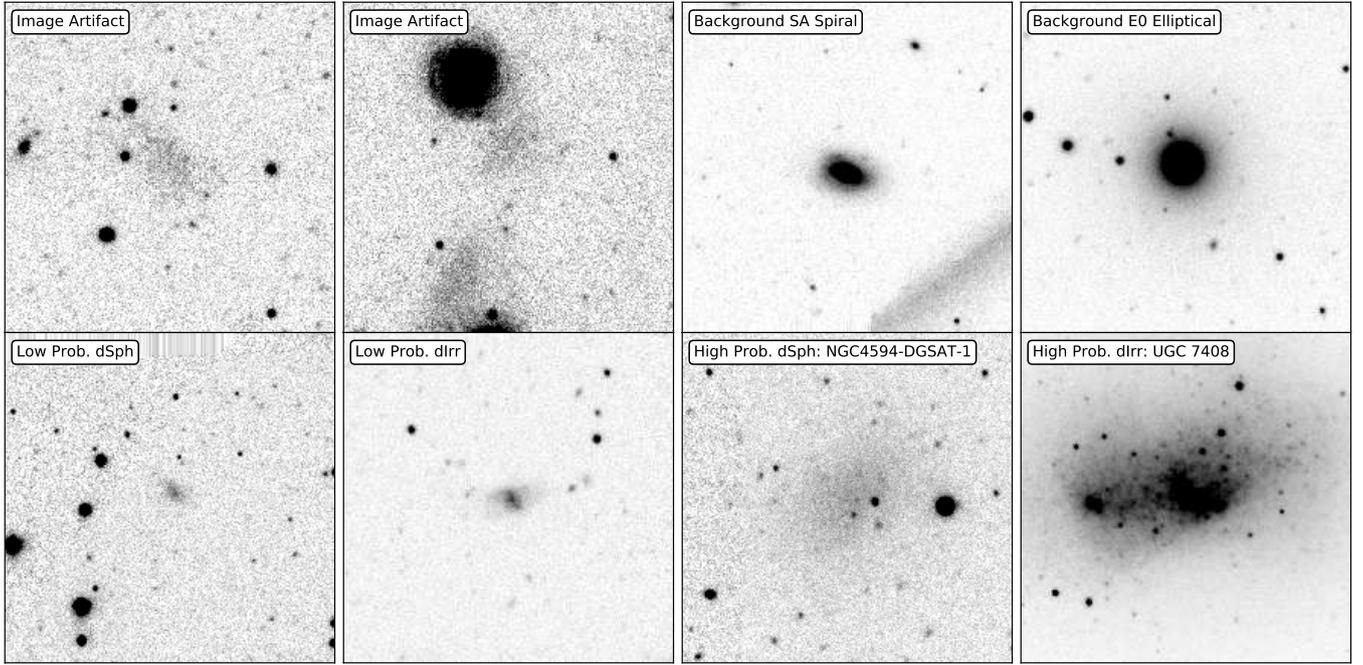}
	\caption{Classifications of example objects using the scheme described in \ref{visual_det}.}
	\label{fig:eg_objs}
\end{figure*}

We scanned for nebulous, diffuse and hazy objects with brightness profiles reminiscent of dwarf elliptical galaxies, while also considering brighter, more irregularly shaped galaxies that may be irregular and blue compact dwarf galaxies. We considered qualitative characteristics of dwarf galaxies including apparent magnitude, luminosity profile, galactic morphology and physical size based on prior experience to register candidate detections. It is however, still inherently difficult to differentiate background spiral galaxies or interacting galaxies with irregular dwarf galaxies, as PSF smoothing of detailed morphology can produce features that could be interpreted as belonging to either type of galaxy in the case the spiral galaxy is distant (and angular size matching the seeing). Compact dwarfs such as Blue Compact Dwarfs (BCDs) or Ultra Compact Dwarfs (UCDs) can lose semblance of any features and become indistinguishable from foreground stars, as their half-light radius (as low as a few 10s of parsecs) can approach the seeing limit. Our conservative approach therefore means we discount the majority of small angular size irregular or compact candidates. 
Finally, careful consideration is given to the object's surroundings, as image artifacts introduced by nearby bright objects as a result of the telescope optics or the detector itself can produce image ghosts and artifacts which resemble low-surface brightness galaxies. Each candidate we present here are assigned a \textit{high} or \textit{low} probability of being a real satellite galaxy of the NGC2683 system, as shown in Table \ref{tab:photometry}. This probability is determined on a case-by-case basis using the considerations above, which we summarise here. A high probability dwarf satellite must satisfy the following:
\begin{itemize}
  \item Lack characteristic morphology of giant galaxies; spiral arms or \textit{cuspy} cores.
  \item Exhibits expected surface-brightness vs. apparent magnitude ratios, as in Fig. \ref{fig:parameter_space}
  \item Has half-light radius $>\,\sim6$ arcseconds, or 300\,pc at NGC2683.
  \item Displays symmetric morphology with smooth cores in the case of suspected dE or dSph candidates.
  \item Displays asymmetric morphology in the case of dIrr candidates.
  \item Be absent of nearby image artifacts.
  \item The object is visible in comparable alternative surveys (only if such a survey exists).
\end{itemize}
A low probability object generally fails at least one of these conditions. For an example of this classification scheme applied to sample objects, see Fig. \ref{fig:eg_objs}. Given membership probability is based on a number of factors that are difficult to quantify and that can be influenced by the authors, we chose this qualitative approach as opposed to poorly defined and arbitrary numerical measurements of probability. If numerical probabilities are preferred, then the assumption that \textit{high} probability candidates have a 90 percent chance and \textit{low} probability candidates a 50 percent chance of belonging the NGC2683 system is reasonable.

We also referenced the SIMBAD and NASA/IPAC and databases to filter out candidates, which may be known background or foreground galaxies within the survey footprint that have measured line-of-sight (LOS) velocities. Comparisons of the LOS velocity of an object with the LOS velocity of NGC2683 can discriminate members of the system. For the brighter potential candidates, we cross-referenced them in SDSS and HSCLA2016 \citep{HSC_leg} images for confirmation, which played a crucial role in filtering out candidates that were image artifacts in our data. Additionally, no foreground galaxies were identified in the footprint and the nearest background galaxies with known LOS velocities as shown in Fig.\ref{fig:footprint} are not near enough for us to consider which host our candidates are bound to.

\subsection{Photometric modelling} \label{photometric_modelling}
As part of our modelling process, we use \textsc{galfit} 3.0.7 \citep{2010AAS...21522909P} to measure structural parameters for our candidate galaxies. \textsc{galfit} uses parametric functions to model objects in two-dimensional digital images, and thus is capable of modelling galaxies, stars, globular clusters, stellar disks and a variety of other astrophysical objects, though we mainly use a subset of this functionality focused on the S\'ersic profile \citep{Sersic1963}. The S\'ersic power law defined used by \textsc{galfit} is:
\begin{equation}
	\Sigma(r) = \Sigma_{\text{eff}} \exp{\left[-\kappa \left(\left(\frac{r}{r_e}\right)^{\frac{1}{n}} -1 \right)\right]}
\end{equation}
where $r_{\text{e}}$ is the effective radius that contains half the total flux, $\Sigma_{\text{eff}}$ is the pixel surface brightness at the effective radius, $n$ is the S\'ersic index or concentration parameter and $\kappa$ is a dependent parameter coupled to $n$ to ensure the integrity of the effective radius.

For each of our satellite candidates, we use \textsc{galfit} to generate a best-fitting 2D light profile and a residual image where the model has been subtracted from the original exposure. The residual image is a direct illustration of the accuracy of the model: if the model exactly fits the galaxy, the residual image should show no trace of the candidate galaxy and resemble the sky background. For dwarf elliptical or dwarf spheroidal {dE/dSph} type galaxies the appropriate representation is a single component S\'ersic fit. We label dwarf galaxies with smooth morphology and $M_{g}>-16$ as dSph galaxies and $M_{g}<-16$ as dE galaxies \citep{Grebel2003}. Galaxies that are not well approximated by a single S\'ersic model are generally transitional or irregular dwarf galaxies ({dT/Irr}), where dust and star forming regions or tidal perturbations lead to asymmetric structures in their light distribution. Consequently, dwarf spheroidal/elliptical galaxies are generally well described by a single symmetric Sérsic profile with a typical shape parameter $0.5<n<1.3$ \citep{Poulain_2021}. The structural parameters of these candidate galaxies are the parameters of the best fit S\'ersic model. For all candidates, we assume the distance to them is equal to the distance to NGC2683.
For irregular galaxies with more complex morphologies fitting multiple overlapping Sérsic components is often necessary to best describe the satellite galaxy. Overall, this process was only used to model KK69, which ended up being the only galaxy present in our data that displayed irregular morphology. This iterative approach for irregular morphology is described in greater detail in Appendix \ref{appendix_a}.

\subsection{Candidate Descriptions}
\begin{figure*}
	\centering
	\includegraphics[draft=false,width=12cm]{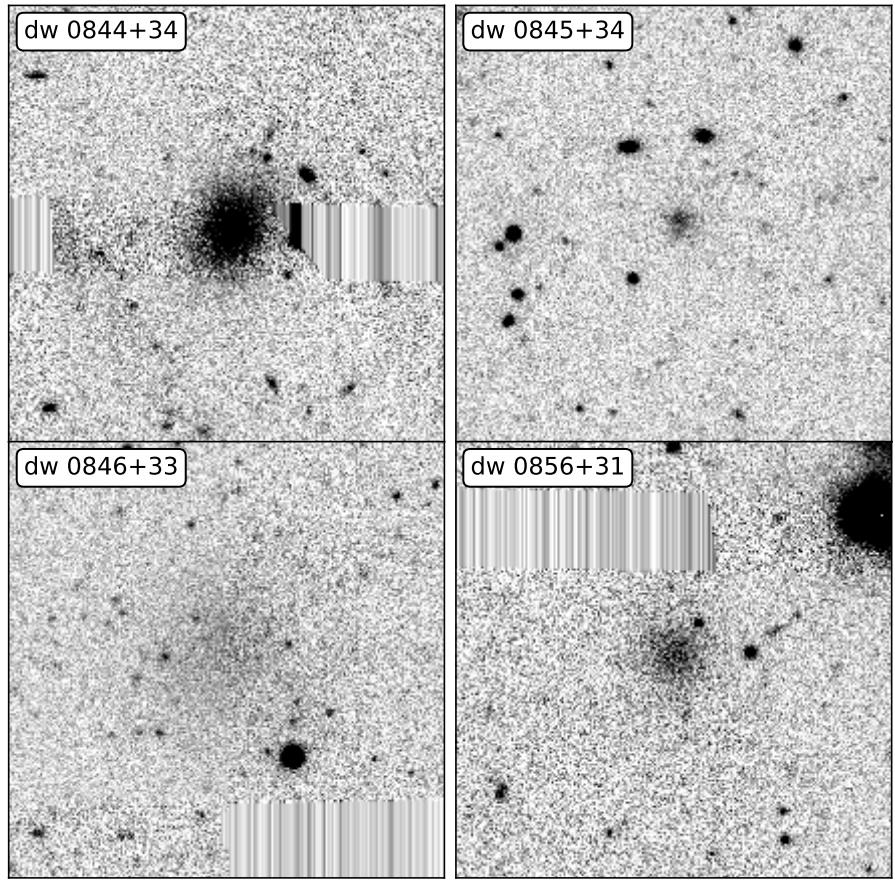}
	\caption{HSC-g images of the four newly detected satellite galaxy candidates within the NGC2683 group. North is up and east to the left. Each image has a side length of $100.8\farcs$ or $~4.6\,$kpc at the distance of NGC2683. 'Barcode' type anomalies are a result of of missing information due to CCD gaps.}
	\label{fig:catalogue}
\end{figure*}

As a result of this process, we describe newly found satellite candidates below. We provide individual descriptions of our candidates and justify the membership probability for each as reported in Table \ref{tab:photometry}. See Figure \ref{fig:catalogue} for raw images of our candidates where each of them are displayed in order, left to right, top to bottom.

\subsubsection{dw 0844+34}
This is an archetypal dwarf spheroidal galaxy given its morphology and model fit parameters and highly likely at least nearby to NGC2683. However this satellite's distance from NGC2683 is large and places it beyond the virial radius of NGC2683. It's early-type morphology indicates it likely resides in a high density environment or bound to a host, as opposed to the field \citep{FOGGIE}. Given that the nearest host galaxy other than NGC2683 in the field of view (NGC2649) resides far behind NGC2683, we argue it is more likely a satellite of NGC2683, thus we assign it a high membership probability.

\subsubsection{dw 0845+34}
A small diffuse object that is possibly a dwarf spheroidal. It's angular size however is only slightly larger than the seeing limit and thus if it were a background galaxy we would expect any structure, upon which background object discrimination is primarily based, to be smoothed by the PSF. This object's surface brightness is also on the high end for a dwarf spheroidal of that size ($\sim170\,pc$, see Fig.\ref{fig:parameter_space}), but is comparable to a few galaxies in the Local Volume, in particular the Sculptor dwarf galaxy \citep{Irwin1995}. Additionally, it lies well beyond the virial radius of NGC2683 indicating this satellite galaxy might not be bound to NGC2683. We have determined this object is worthy of consideration however, but assign it as a low membership probability.

\subsubsection{dw 0846+33 or dw0846p3348}
This is a large, low-surface brightness dwarf spheroidal galaxy that is highly likely a member of the NGC2683 system given it's morphology. This galaxy is unusually diffuse for it's total luminosity, based on its position in the $\mu_e - M_g$ plot as shown in Fig.\ref{fig:parameter_space} in comparison to LV galaxies, with a low surface brightness near the detection limit. This indicates this object may be better classified as an ultra diffuse galaxy (UDG) as its central surface brightness is $\mu_{g,0}=26.2\,\text{arcsec}^{-2}$, which is far below the general cut-off of $\mu_{0}>24\,\text{arcsec}^{-2}$ \citep{van_Dokkum_2015}. With no reason to doubt its candidacy, we assign a high membership probability. This object was independently discovered by \citealt{Carlsten_2022} and named dw0846p3348.

\subsubsection{dw 0856+31 or dw0856p3155}
This archetypal dwarf spheroidal is highly likely a member of the NGC2683 system given its morphology, angular size and surface brightness, as quenched satellites are expected in host galaxy environments \citep{FOGGIE}, despite this object residing just outside of the virial radius. We assign it a high membership probability. This object was independently discovered by \citealt{Carlsten_2022} and named dw0856p3155.

\subsection{Survey Completeness} \label{sec_completeness}
\begin{figure}
	\centering
	\includegraphics[draft=false,width=8cm]{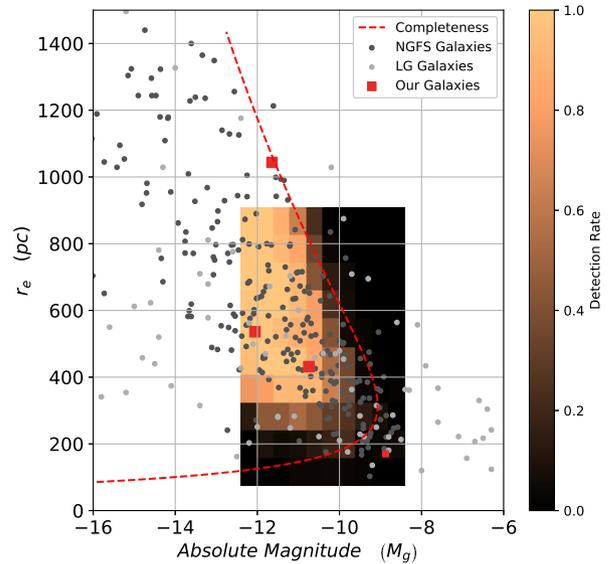}
	\caption{Survey completeness illustrated in the $r_e - M_g$ plane. Galaxies from the Fornax cluster \citep{Eigenthaler2018} and the Local Volume \citep{Karachentsev_2013} are plotted are grey dots. The red squares are satellite candidates presented in this paper. Both the Analytical relation as a red dashed line, and the 2D histogram formed from finding generated galaxies by-eye demonstrate the completeness.}
	\label{fig:completeness}
\end{figure}

The completeness limits of our survey in terms of total magnitude, surface brightness and half-light radius is shown in Figure \ref{fig:completeness}. We use the completeness relation from \cite{Ferguson_Sandage_1988} to plot the completeness function.

\begin{equation}
	m_{\text{tot}} = \mu_{\text{lim}} - \frac{r_{\text{lim}}}
	{0.5487 r_{e}}-2.5*\text{log}\left[2\pi \left(0.5958 r_{e}\right)^2\right]
\end{equation}

We set $\mu_{\text{lim}}=26$ and $r_{\text{lim}}=1.9$ arc-seconds which roughly corresponds to the seeing at the time of observation, produces a good fit and matches what we expect given by-eye analysis of the images. Additionally, we placed 4000 random simulated dwarf elliptical galaxies which were overlaid on top of a sample of the same telescope images one-by-one. This 'sample' consists of underlying fields that don't contain significant contamination from foreground stars, image artifacts or contain known satellite candidates. These simulated dSph galaxies have random parameters in the range of $100\,$pc$\,\leq r_{\text{e}} \leq900\,$pc, $-12.5\leq M_{\text{g}} \leq-8.5$. The axis ratio and S\'ersic index are allowed to vary from $0.5-1.0$. We used the same techniques used to discriminate candidates to determine if the simulated dSph was detected in the image. These results form a 2D histogram which is underlaid in Fig.\ref{fig:completeness}. For $r_{\text{e}}>400\,$pc, we find that the \cite{Ferguson_Sandage_1988} relation follows the decreasing detection rate gradient well, but for $r_{\text{e}}<400\,$pc this is not the case. The analytical relation does not take into account factors relevant to this analysis for detecting dwarf galaxy candidates, including discrimination of background elliptical and spiral galaxies, image artifacts, or that smaller objects are more likely to be completely obscured by bright foreground and background objects. The histogram encapsulates both detection limits of the images and criteria used to find candidates by-eye. Ultimately, this means that though compact and bright objects are readily visible, they don't register as a satellite candidate in this scheme and resemble a foreground object. This results in the departure of the relation from the histogram for $r_{\text{e}}<400$\,pc and the real chance that compact dwarf galaxies, including BCDs or UCDs go undetected. For future reference, our survey is 100 percent complete for $M_{\text{g}}<-10$, and 50 percent complete for $-10<M_{\text{g}}<-9$, excluding compact objects of $r_{\text{e}}<300$\,pc.

Otherwise for archetypal dEs, dSphs and Irr galaxies reminiscent of those in the Local Group, the completeness plot demonstrates our survey is adequately complete down to $M_{\text{g}}\sim-10$. This allows us to make fair comparisons with satellite galaxy systems of at least the Local Group, though it has recently been suggested that there exists a significant number of mis-categorised compact dwarfs in the Centaurus A/M83 group \citep{Dumont2022} and thus comparisons to richer, denser galaxy environments may neglect contributions from compact satellite galaxies.

\begin{figure}
	\centering
	\includegraphics[draft=false,width=8cm]{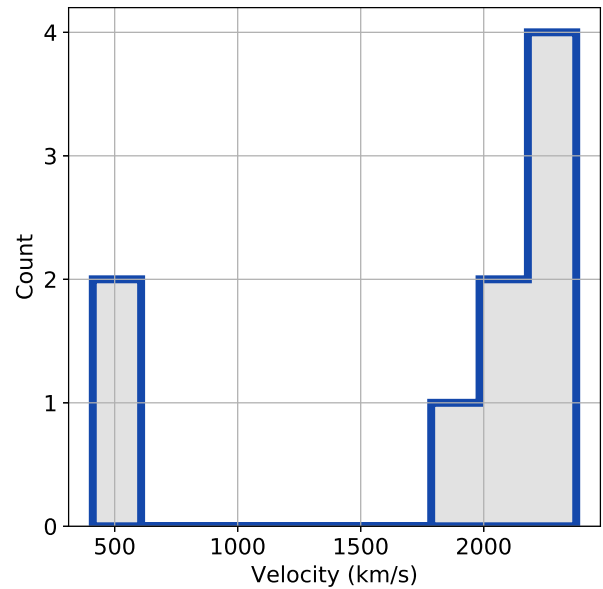}
	\caption{Histogram of all known recessional velocities of galaxies within the survey footprint out to 2500\,km\,s$^{-1}$. Data is extracted from the NASA/IPAC Extragalactic Database and objects with recessional velocities $<200$\,km\,s$^{-1}$ are removed as stars within the Milky Way are often miscategorised as galaxies in the SDSS and WISEA catalogs.}
	\label{fig:footprint}
\end{figure}

Finally, we also introduce a histogram of the known recessional velocities of galaxies within the survey footprint (as seen in Fig. \ref{fig:NGC2683_coverage}) in Fig. \ref{fig:footprint} out to $2500$\,km\,s$^{-1}$. Both NGC2683 and KK69 are present in the $500$\,km\,$^{-1}$ bin and nearest galaxies behind the NGC2683 environment are concentrated at 2000 \,km\,$^{-1}$. The NGC2683 Group is then expected to be well separated from other galaxies in the footprint by at least 30 Mpc (using velocity as a proxy for distance), and thus the satellite candidates we present here are highly likely members of the NGC2683 system.

\section{Discussion}
\subsection{The $\mu_e - M_g$ and $r_e - M_g$ plane}
We use the $\mu_e - M_g$ and $r_e - M_g$ parameter spaces as tools to test the membership of these galaxies in the NGC2683 system. For galaxies of all morphological types, these parameters are correlated \citep{Kormendy1974, Caldern2020} and this correlation is used to discriminate the candidacy of newly detected galaxies (as in \cite{Muller_2016}). We compare the photometry of our sample with photometry of galaxies in the Fornax cluster, from the Next Generation Fornax Survey (NGFS) \citep{Eigenthaler2018} and Local Volume (LV) galaxies from the Catalog and Atlas of the LV Galaxies \citep{Karachentsev_2013}, a maintained and updated catalog of galaxies within the LV. These surveys report magnitudes in different bands, so we used the SSDS band conversions \citep{Jester_2005}, average $B-V$ colours for the generally blue irregular dwarf galaxies \citep{Makarova_2009} and $B-V$ colours for generally red elliptical galaxies \citep{Chiosi_Carraro_2002} to convert the $B$ and $V$ magnitudes into approximate \textit{HSC-g} band magnitudes. Given we present only the \textit{HSC-g} band photometry due to some poor weather at the time of observing, we are required to estimate the colour in this manner. The error in these estimates is high (up to $\sim0.8$ mag in $m_g$), but ultimately the comparison with our candidates to known dwarf galaxies is largely qualitative and low uncertainties are not largely beneficial.

\begin{figure*}
	\centering
	\includegraphics[draft=false,width=16cm]{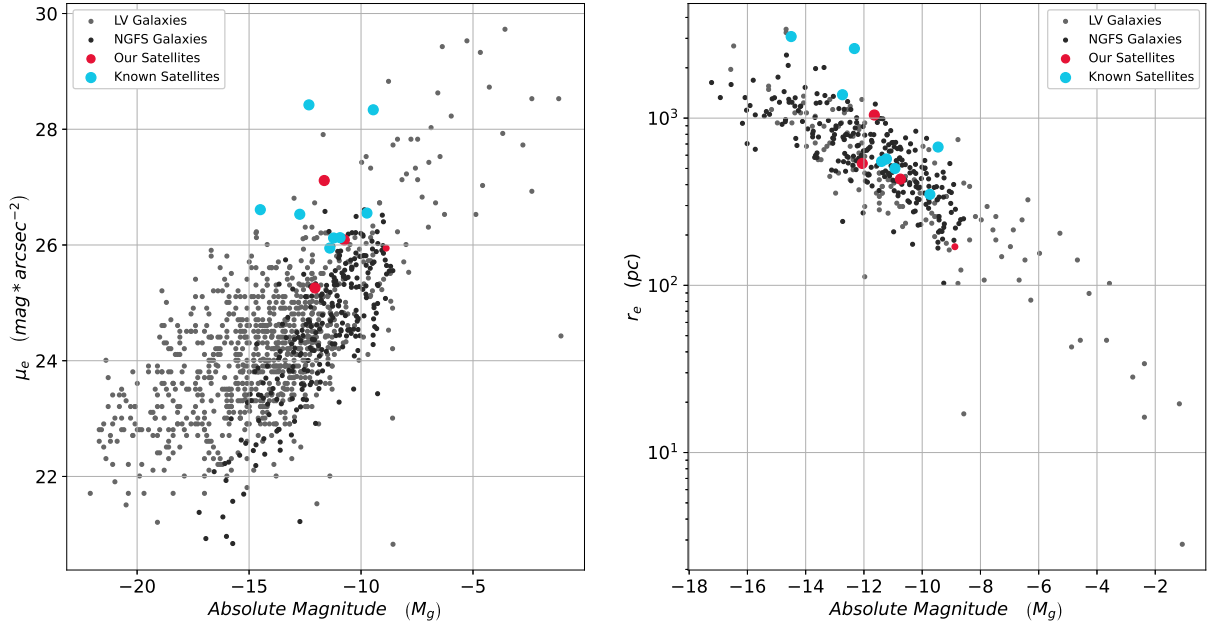}
	\caption{\textbf{Left panel:} The $\mu_e - M_g$ parameter space of galaxies in our sample, the NGFS and Local Volume satellite galaxies. 
	\textbf{Right panel:} The $r_e - M_g$ parameter space of galaxies in our sample, the NGFS and Local Volume satellite galaxies.
	In both panels the blue circles represent our satellite galaxy candidates and red circle represent candidates and confirmed satellites already known. The size of the circles are scaled with membership probability.}
	\label{fig:parameter_space}
\end{figure*}

The results, as shown in Figure \ref{fig:parameter_space}, fail to strongly exclude any particular galaxy from our sample and in fact reinforces their position as probable real satellite galaxies. We also plot the positions of the population of known satellites, which fit into the locale of the parameter space as expected.

\subsection{The Satellite Plane of NGC2683} \label{NGC2683_sat_plane}
Visual inspection of the satellite galaxy distribution as shown in Fig.\ref{fig:NGC2683_coverage} reveals what appears to be an elongated, somewhat asymmetric 2D arrangement of satellite galaxies approximately aligned with the polar axis (a separation of 28 degrees) of NGC2683, which resembles similar structures of satellite galaxies found in the Milky Way \citep{Pawlowski2012} and M31 \citep{Ibata2013,Conn2013} systems. To study this elongation in more detail, we use Principle Component Analysis (PCA) to determine the axis ratio and position angle of the best fitting ellipse about the distribution of satellite galaxies. We use the eigenvectors ($\textbf{v}_{1}$ and $\textbf{v}_{2}$) and the eigenvalues ($\lambda_{1}$ and $\lambda_{2}$) of the covariance matrix formed in the parameter space consisting of the 2D projected coordinate of each satellite galaxy. 
These eigenvectors are an orthonormal basis which can be interpreted as the semi-minor and semi-major axis of an ellipse in the right-ascension and declination parameter space. We can extract the most probable axis ratio of the distribution of the satellite galaxies in the projected 2D space from the eigenvectors forming this ellipse. In that case, the semi-major and semi-minor axes are:
$a = \sqrt{\lambda_{1}}$, $b = \sqrt{\lambda_{2}}$.
The length of minor and major axis are calculated to be $0.37$ and $1.44$ respectively, and the axis ratio is $b/a=0.26$ as plotted in Figure \ref{fig:NGC2683_coverage}. The uncertainty of the measured axis ratio can be approximated with bootstrapping; we select 10 random candidates allowing the same candidate to be selected more than once and measure the axis ratio. This is repeated $10^5$ times and the axis ratios for a Gaussian distribution of $\langle b/a\rangle\approx0.68\pm0.16$. The peak of this distribution lies $2.5\,\sigma$ away from the actual measured value due to a number of satellites contributing to the flattening in the system as seen in Figure \ref{fig:NGC2683_coverage}. We adopt $\pm0.16$ as the uncertainty in the measured axis ratio. Thus the observed axis ratio is $0.26\pm0.16$. As shown in Figure. \ref{fig:NGC2683_coverage}, the major axis of the PCA ellipse of NGC2683 also approximately aligns with the polar axis of NGC2683. 

Using a similar analysis, \cite{Heesters2021} noted when a flattened structure was detected around an early-type host in a large sample of MATLAS satellite galaxy systems, the axis ratio spanned the range of $0.1-0.5$ and appeared to be approximately aligned with the polar axis. The parameters of the flattened structure of NGC2683 compare well to these galaxies with detected flattened structures.

This axis ratio without context offers little meaning. We used a Monte-Carlo approach to generate random, spherically symmetric distributions of satellite galaxy positions in a 3D sphere out to the virial radius of NGC2683 with an underlying flat radial distribution. We simulated the distribution of 12 satellite galaxies and determined the axis ratio using the same PCA axis ratio fitting approach $10,000$ times. This approach generates a Poisson distribution of axis ratios which can be approximated as a Gaussian distribution with an axis ratio of $\approx0.72\pm0.14$ for an underlying isotropic distribution of 12 satellite galaxies. The peak of the measured axis ratio of $\approx0.26$ lies within the top <1 percent of this distribution. This 'isotropic' axis ratio approaches a value of 1 when the number of satellites increases. 

\begin{figure*}
	\centering
	\includegraphics[draft=false,width=15cm]{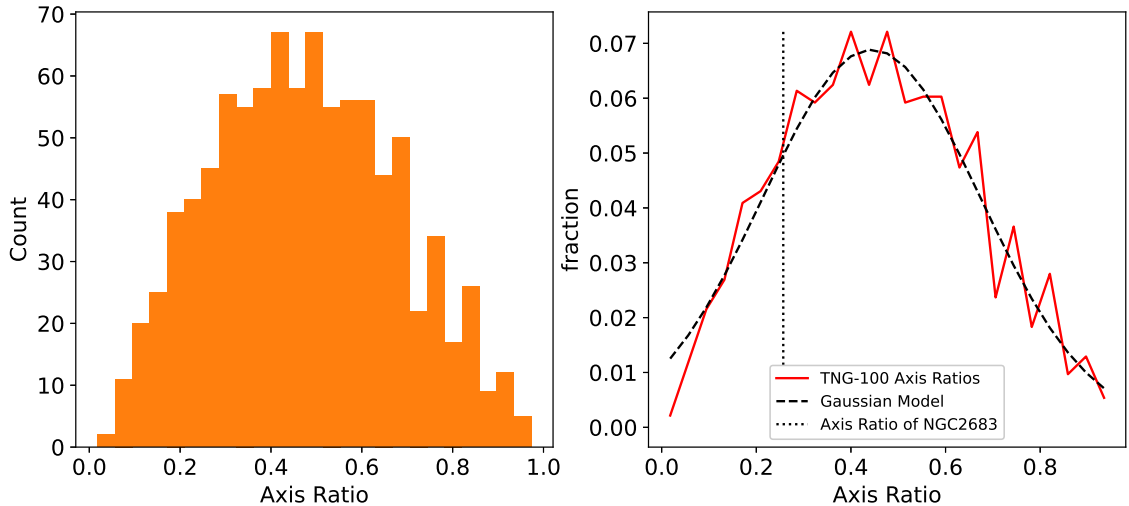}
	\caption{\textbf{Left panel:} A histogram of the axis ratios of the satellite systems from the selected comparison galaxies in the \emph{TNG100-1} model at the $z=0$ slice \citep{TNG100}. The selected galaxies are $\approx2\,$Mpc from the nearest $L^*$ host galaxy, have total mass ($\approx0.5-5.0\times\,{10^{12}\,M_\odot}$), and at least 12 satellite galaxies within 2 times the $R_{200}$ radius. Only the 12 brightest satellites were selected if the host possesses >12 satellites within this radius.
	\textbf{Right panel:} The same histogram as in the left panel drawn as a line graph (red solid line), with the best-fit Gaussian distribution (black dashed line), compared to the axis ratio of the NGC2683 system (vertical black dotted line). $\sim19$ percent of simulated systems lie below the vertical dotted line.
	}
	\label{fig:NGC2683_sim_comp}
\end{figure*}

We additionally compare the distribution of satellite galaxies in the NGC2683 system to recent cosmological simulations. We use the IllustrisTNG \emph{TNG100-1} cosmological, gravo-magnetohydrodynamical model \citep{TNG100} to compare the NGC2683 system to a simulated host galaxy environment. This \emph{TNG100-1} model consists of a simulated box-shaped volume with a side length of 110.7 Mpc, using adaptive resolution elements to fully capture both non-baryonic and baryonic physics on the scales of star-forming regions, black holes, galaxies and galaxy clusters from redshift $z\approx20$ to $z=0$. We examine the two-dimensional projected satellite galaxy distributions of 929 simulated host galaxy environments within the \emph{TNG100-1} model at the time slice $z=0$. Host halos are identified using Friends-of-Friends (FoF) algorithm and satellites through the Subfind algorithm. From this model, we selected isolated galaxy groups which at the time of the snapshot, are at least $\sim1.5\,$Mpc away from the nearest $L^*$ galaxy group to emulate the isolation of NGC2683. We additionally restricted the analysis to Milky Way-like host galaxies in terms of total mass ($\approx 5-50\times {10^{11}\,M_\odot}$) and those that had at least 12 sub-halos within the two times the virial radius ($R_{200}$) with $M_{g}<-6$, to exclude dark halos. If it possessed more than 12 sub-halos, the 12 brightest sub-halos were selected. Again we repeat a similar analysis; we view these simulated host galaxies and their satellites from a random orientation along an axis and project the selected sub halos onto the 2D perpendicular 2D plane, then we use PCA to find the axis ratio of this distribution. As shown in Figure \ref{fig:NGC2683_sim_comp}, we collect the measured axis ratios into a histogram and fit a Gaussian distribution, this generates a 2D axis ratio distribution with a mean and error of $0.44\pm0.23$ (taken from the parameters of the fit Gaussian distribution) which reveals that $\approx19$ percent of the 929 selected simulated systems are more flattened than the NGC2683 system. 

\subsection{Galaxy Luminosity Function of the NGC2683 environment}

\begin{figure*}
	\centering
	\includegraphics[draft=false,width=15cm]{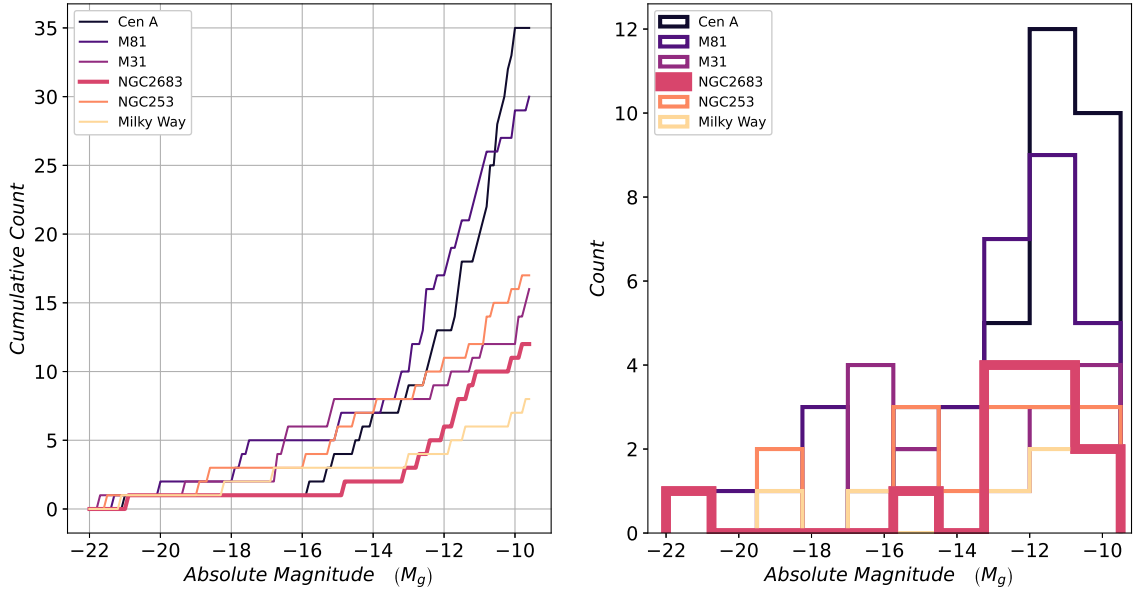}
	\caption{\textbf{Left panel:}The cumulative Galaxy Luminosity Functions (GLF) of the Centaurus A, M81, NGC253, M31, Milky Way and NGC2683 systems truncated at $M_{g}=-9.5$, the approximate magnitude limit of our survey of the NGC2683 system. \textbf{Right panel:} The Luminosity function of the NGC2683 system and the same comparison systems.}
	\label{fig:GLF}
\end{figure*}

One useful tool to study the nature of the system of satellites in the NGC2683 system is to compare the luminosity function to other well-studied systems. Primarily, it can illustrate the \textit{TBTF} problem, but in more detail also highlights differences or similarities in the sub-halo mass functions. In the left panel of Fig. \ref{fig:GLF} we present the cumulative Galaxy Luminosity Functions (GLFs) of the Centaurus A, M81, NGC253, Andromeda (M31), Milky Way and NGC2683 systems \citep{Karachentsev_2013}. The galaxies presented in this graph are those assumed to be a satellite galaxy of the labelled host, within a projected distance of 400\,kpc. This is not perfect, particularly for NGC253 where dwarf galaxies belonging to the group appear to be near NGC253 in the projected plane, but have separations along the line of sight of $>400\,$kpc, much more than the estimated virial radius of $186\,$kpc \citep{MartnezDelgado2021}. However, we argue that including dwarf galaxies beyond the virial radius is not unreasonable given indications that planar alignments of dwarf galaxies appear to extend to scales of $1-2\,$Mpc \citep{Pawlowski2013_b} and that satellites beyond the virial radius are not necessarily unassociated or unbound from their hosts.

Qualitatively, the luminosity functions follow the same profile for all the provided galaxies, but are scaled by a factor and fixed at the point of the host galaxy luminosity, which is consistently $M_B \approx-21 $. From this we conclude that the NGC2683 system consists of an average number of satellite galaxies when compared to similar galaxy environments, namely, NGC253 and the Milky Way. M31, M81 and Cen A are larger galaxies with more vigorous accretion histories that naturally lead to higher satellite galaxy counts. The NGC2683 system includes marginal, unconfirmed candidates which may be discounted with further analysis and thus fall in-line with the Milky Way. Additionally, the brightest NGC2683 satellite, KK[98a]69, has a luminosity of $M_g\approx-14.5$, which is seven magnitudes fainter than NGC2683 and is dim compared to the other host galaxy environments shown in Fig.\ref{fig:GLF}.

This large luminosity gap between the host and brightest satellite is known to be larger in comparable environments in simulations \citep{Bullock2017} and is a manifestation of the Too-Big-To-Fail (TBTF) problem. Simulations appear to produce greater numbers of brighter satellite companion galaxies than observed in the local universe, which on average produces a smaller magnitude gap between the brightest satellite galaxy and its host. Here we seek to test this problem.

\begin{figure}
	\centering
	\includegraphics[draft=false,width=8cm]{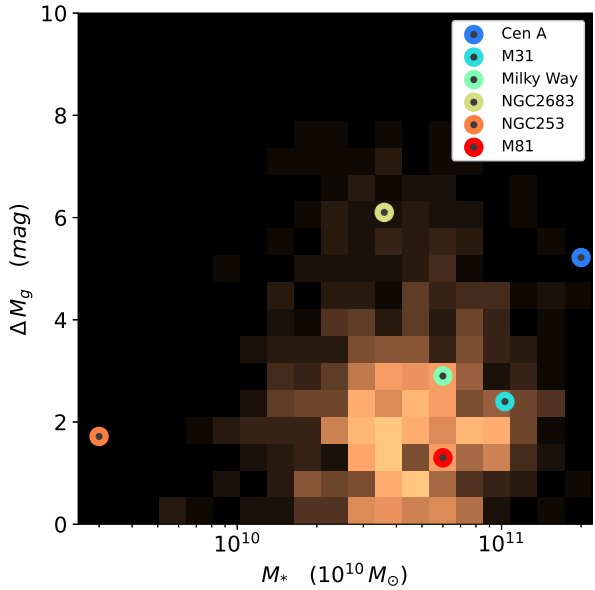}
	\caption{The galaxy luminosity function magnitude gap versus host galaxy stellar mass. The 2D histogram is formed from the converted g-band magnitude gap between the host galaxy and the brightest satellite within 400 kpc ($\Delta\,M_{g}$) and the Host galaxy stellar mass ($M_{*}$) of \emph{TNG100-1} simulated environments as described in Section \ref{NGC2683_sat_plane}. Overlaid on this histogram are data points of sample galaxies as observed. The stellar masses for the observed environments are drawn from a variety of sources as follows: Cen A \citep{Wang_2020}, M31 \citep{Sick_2014}, Milky Way \citep{Licquia_2015}, NGC2683 \citep{Vollmer_2016}, NGC253 \citep{Bailin_2011} and M81 \citep{Sheth_2010}.
	}
	\label{fig:NGC2683_maggap_hist}
\end{figure}

To do so, we directly compare the magnitude gaps between the brightest satellite galaxy for the reference galaxies, against the same sample of galaxy environments from the \emph{TNG100-1} model as identified in section \ref{NGC2683_sat_plane}. We plot the galaxy luminosity function magnitude gap against the host galaxy stellar mass, as shown in Figure \ref{fig:NGC2683_maggap_hist}.

We find from this graph that NGC2683 has a magnitude gap of $6.1\,m_{g}$, which is greater than all of the comparison Local Volume environments and 95\% of the \emph{TNG100-1} simulated environments. The comparison environments M31, M81, NGC253 and the Milky Way possess luminous companions with total absolute magnitudes in the range of -16 to -19 that are gravitationally bound to their hosts. This is consistent with simulations based on visual inspect of the graph. The brightest satellites for these reference environments are as follows: M33 for M31, M82 for M81, NGC247 for NGC253 and the LMC for the Milky Way. For M33 and NGC247 in particular, they reside outside of the virial radii of their hosts, but with distances and velocities relative to their hosts that make them plausible satellites. However, NGC2683 does not possess a luminous companion comparable to these satellites, which is uncommon in simulations, only occurring in about 5\% of simulated systems. Overall we do not find any meaningful differences between simulated environments and the observed environments based on this analysis. There is also no obvious correlation between the host galaxy stellar mass and the magnitude gap. A more thorough and robust analysis fitting an analytical function such as a Schechter function \citep{Schechter_1976} is challenging given the small number of satellites in the NGC2683 system, the absence of luminous satellites and the completeness limits of our observations. A more in-depth study with a larger sample of environments may be helpful in revealing the fundamental processes, if they exist, that produce large magnitude gaps.

\begin{figure*}
	\centering
	\includegraphics[draft=false,width=15cm]{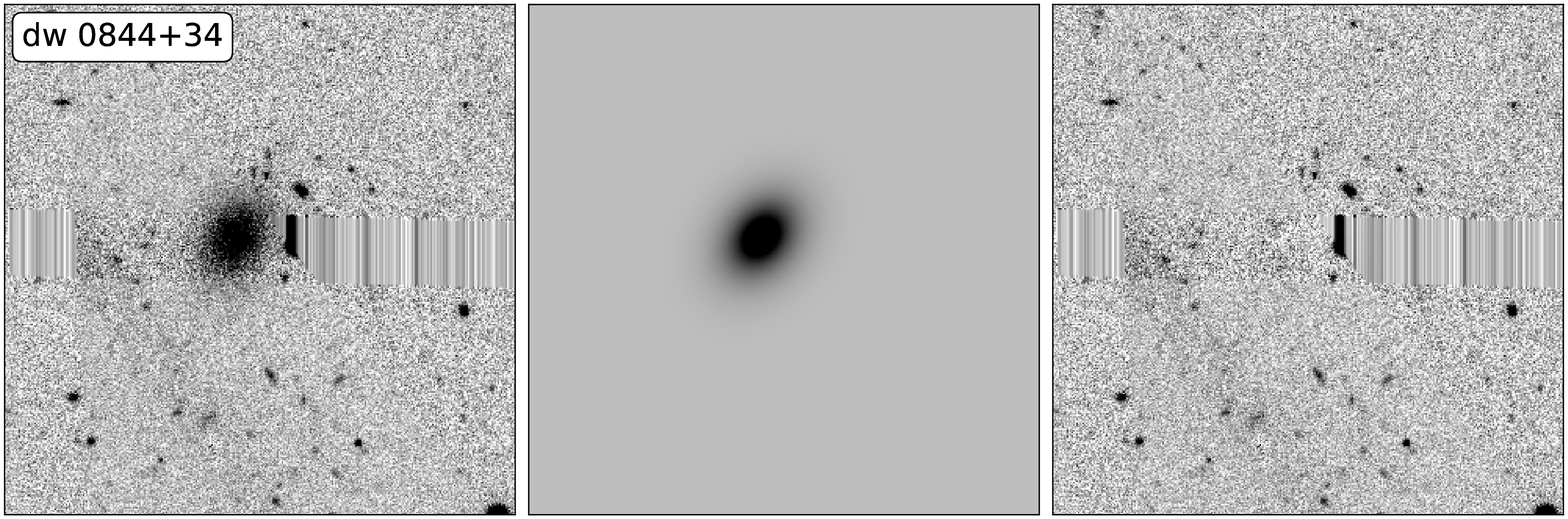}
	\includegraphics[draft=false,width=15cm]{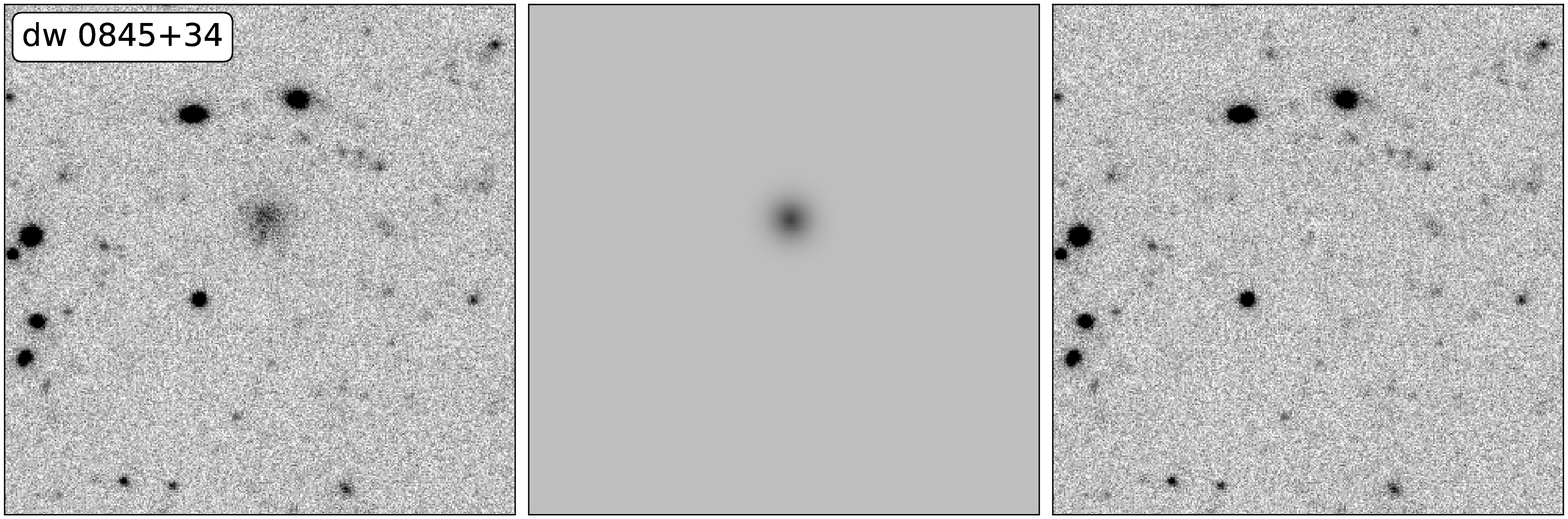}
	\includegraphics[draft=false,width=15cm]{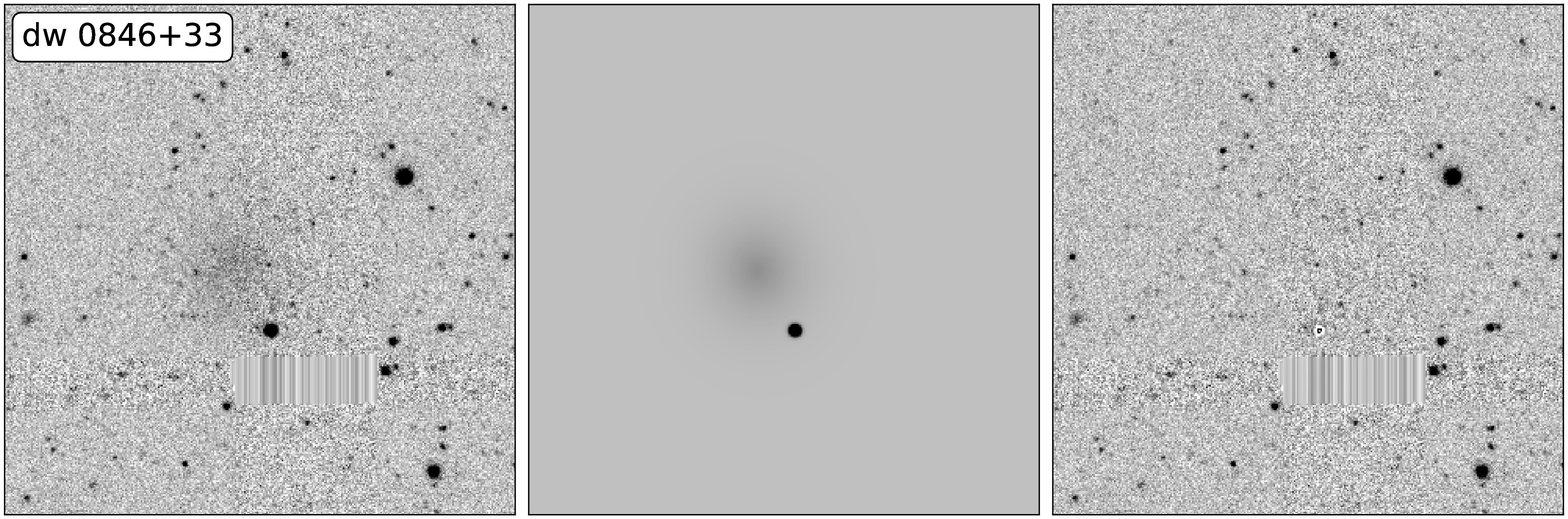}
	\includegraphics[draft=false,width=15cm]{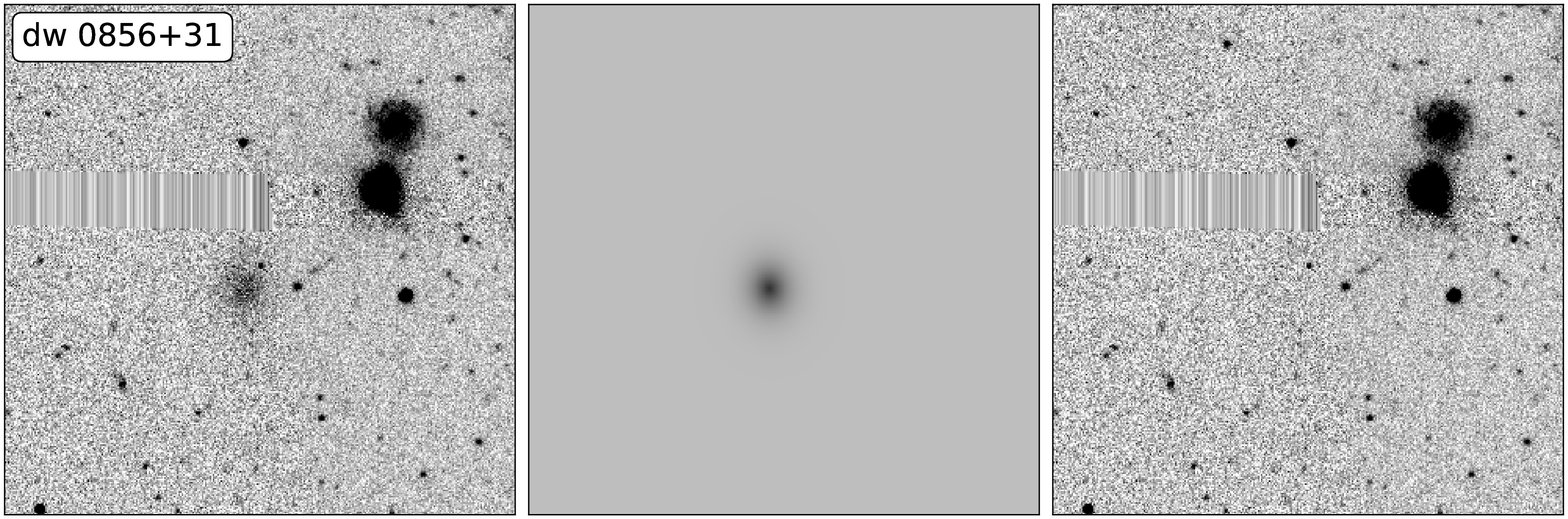}

\caption{\textsc{galfit} results for our dwarf galaxy candidates detected in the NGC2683 survey area. Each row is divided into three images: (left) the main coadded HSC-g exposure; (middle) the \textsc{galfit} model; (right) residual image obtained by subtracting the model from the main exposure. Model parameters are listed in Table \ref{tab:photometry}.}
\label{fig:photometry_images}
\end{figure*}

\begin{table*}
\caption{Photometry, structure parameters and spatial information of all satellites and candidates of the NGC2683 System. Photometric properties are derived from the best-fitting \textsc{galfit} models from either our analysis or previous analysis \citep{Javanmardi_2016}. NGC2683-dw1 has no prior \textsc{galfit} photometry, so we use the HSCLA2016 images for this object. In the top partition of the table, we present the previously known candidates of NGC2683 and in the bottom, our new candidates. Below each measurement, we present uncertainties where available. Each column is as follows, (1): Name of candidate, (2): J2000 Right Ascension, (3): J2000 Declination, (4): Galaxy Morphology, (5): total apparent magnitude, (6): Galactic extinction, (7): absolute magnitude, (8): half-light radius in arc seconds, (9): half-light radius in parsec, (10): mean surface brightness within the effective radius, (11): S\'ersic index, (12): axis ratio $b/a$, (13): projected distance from NGC2683, (14): TRGB distance to candidate, (15): systemic velocity, (16): membership probability. For satellites without measured distances, we assume the distance is that of NGC2683, $\mu = 29.86\pm0.06$ mag \citep{Karachentsev_2015_A}. Photometric bands are indicated next to photometric measurements in the table.}
\medskip
\resizebox{\textwidth}{!}{%
\tabcolsep=0.0pt
\begin{tabular}{*{15}{c@{\hspace{1em}}}c} 
	\hline
    ID & R.A. & DEC & Type & $m$ & $A_g$ & $M$ & $r_e$ & $r_e$ & $\langle\mu_e\rangle$ & $n$ & $b/a$ & $\Delta_{N2683}$ & D & $v_{\textit{sys}}$ & Memb. \\
    
    - & hh:mm:ss & hh:mm:ss & - & mag & mag & mag & arcsec & kpc & mag & - & - & kpc & Mpc & km\,s$^{-1}$ & Prob. \\
    
    (1) & (2) & (3) & (4) & (5) & (6) & (7) & (8) & (9) & (10) & (11) & (12) & (13) & (14) & (15) & (16) \\
    
    \hline

   KK[98a]69 & 133.199 & +33.793 & dT/UDG & $15.46$ \textit{HSC-g} & 0.09 & $-14.50$ \textit{HSC-g} & $76.7$ & $2.07$ & $25.66$ \textit{HSC-g} & $0.34$ & $0.83$ & 60.8 & $9.28$ & $464$ & Confirmed \\
   
    - & 8:52:48 & 33:47:35 & - & $\pm0.03$ & - & $\pm0.07$ & $\pm0.4$ & $\pm0.09$ & $\pm0.03$ & $\pm0.01$ & $\pm0.01$ & - & $\pm0.28$ & - & - \\
    
    NGC2683-DGSAT-5 & 133.200 & +32.827 & dSph & $20.08$ \textit{r} & - & $-9.78$ \textit{r} & $6.8$ & $0.31$ & $26.00$ \textit{r} & $0.52$ & $0.55$ & 97.3 & - & - & High \\
   
   - & 8:52:48 & 32:49:37 & - & $\pm0.08$ & - & $\pm0.1$ & $\pm0.4$ & $\pm0.01$ & $\pm0.08$ & $\pm0.05$ & $\pm0.03$ &  & - & - & - \\
   
   NGC2683 dw1 & 133.362 & +33.305 & Irr & $18.53$ \textit{HSC-r} & - & $-11.40$ \textit{HSC-r} & $12.1$ & $0.55$ & $24.5$ \textit{HSC-r} & $0.52$ & $0.52$ & $36.6$ & - & $421$ & Confirmed \\
   
   - & 8:53:27 & 33:18:18 & - & $0.11$ & - & $0.13$ & $0.57$ & $0.02$ & $0.12$ & $0.03$ & $0.01$ & - & - & - & - \\
   
   NGC2683 dw2 & 133.583 & +33.247 & dSph & $18.99$ \textit{r} & - & $-10.87$ \textit{r} & $9.1$ & $0.4$ & $26.48$ \textit{r} & $0.70$ & $0.98$ & 73.0 & - & - & High \\
   
   - & 8:54:20 & 33:14:49 & - & $\pm0.23$ & - & $\pm0.24$ & $\pm2.1$ & $\pm0.1$ & $\pm0.23$ & $\pm0.19$ & $\pm0.14$ & - & - & - & - \\
   
   NGC2683-DGSAT-3 & 133.795 & +33.613 & dSph & $18.59$ \textit{r} & - & $-11.27$ \textit{r} & $11.2$ & $0.51$ & $25.52$ \textit{r} & $0.61$ & $0.61$ & 106.6 & - & - & High \\
   
   - & 8:55:11 & 33:36:47 & - & $\pm0.07$& - & $\pm0.09$& $\pm0.1$ & $\pm0.01$ & $\pm0.07$& $\pm0.04$ & $\pm0.02$ & - & - & - & - \\
   
   KK[98a]70 & 133.847 & +33.559 & dSph & $16.11$ \textit{r} & - & $-13.7$ \textit{r} & $27.3$ & $1.24$ & $25.62$ \textit{r} & $0.78$ & $0.79$ & 112.4 & $9.18$ & - & Confirmed \\
   
   - & 8:55:23 & 33:33:32 & - & $\pm0.06$ & - & $\pm0.08$& $\pm0.7$ & $\pm0.03$ & $\pm0.06$ & $\pm0.02$ & $\pm0.01$ & - & $\pm0.30$ & - & - \\

   dw0846p3300 & 131.558 & +33.000 & Irr/UDG & $17.63$ \textit{HSC-g} & 0.10 & $-12.3$ \textit{HSC-g} & $57.4$ & $2.61$ & $28.42$ \textit{HSC-g} & $0.66$ & $0.50$ & 272.6 & - & - & High \\
   
   - & 8:46:12 & 33:00:00 & - & $\pm0.01$ & - & $\pm0.07$ & $\pm1.9$ & $\pm0.08$ & $\pm0.02$ & $\pm0.01$ & $\pm0.01$ & - & - & - & - \\

   dw0848p3226 & 132.166 & +32.445 & dSph & $20.49$ \textit{HSC-g} & 0.09 & $-9.45$ \textit{HSC-g} & $14.8$ & $0.67$ & $28.34$ \textit{HSC-g} & $0.57$ & $0.65$ & 229.1 & - & - & High \\
   
   - & 8:48:39 & 32:26:42 & - & $\pm0.03$ & - & $\pm0.07$ & $\pm0.73$ & $\pm0.03$ & $\pm0.07$ & $\pm0.04$ & $\pm0.02$ & - & - & - & - \\
   
   \hline

   dw 0844+34 & 131.108 & +34.575 & dSph & $17.90$ \textit{HSC-g} & 0.10 & $-12.06$ \textit{HSC-g} & $11.80$ & $0.54$ & $24.20$ \textit{HSC-g} & $0.88$ & $0.76$ & 386.4 & - & - & High \\
   
   - & 8:44:26 & 34:34:30 & - & $\pm0.01$ & - & $\pm0.06$ & $\pm0.11$ & $\pm0.01$ & $\pm0.02$ & $\pm0.01$ & $\pm0.01$ & - & - & - & - \\
   
   dw 0845+34 & 131.296 & +34.408 & dSph & $21.08$ \textit{HSC-g} & 0.10 & $-8.88$ \textit{HSC-g} & $3.74$ & $0.17$ & $25.16$ \textit{HSC-g} & $0.97$ & $0.86$ & 346.3 & - & - & Low \\
   
   - & 8:45:11 & 34:24:29 & - & $\pm0.05$ & - & $\pm0.08$ & $\pm0.13$ & $\pm0.01$ & $\pm0.06$ & $\pm0.02$ & $\pm0.03$ & - & - & - & - \\
  
   dw 0846+33 & 131.630 & +33.810 & dSph & $18.31$ \textit{HSC-g} & 0.09 & $-11.64$ \textit{HSC-g} & $23.0$ & $1.04$ & $26.28$ \textit{HSC-g} & $0.73$ & $0.93$ & 259.8 & - & - & High \\
   
    - & 8:46:31 & 33:48:36 & - & $\pm0.01$ & - & $\pm0.06$ & $\pm0.4$ & $\pm0.04$ & $\pm0.03$ & $\pm0.02$ & $\pm0.01$ & - & - & - & - \\
  
   
  
   dw 0856+31 & 134.108 & +31.917 & dSph & $19.20$ \textit{HSC-g} & 0.09 & $-10.75$ \textit{HSC-g} & $9.51$ & $0.432$ & $25.16$ \textit{HSC-g} & $0.79$ & $0.85$ & 289.6 & - & - & High \\
   
    - & 8:56:26 & 31:55:01 & - & $\pm0.02$ & - & $\pm0.06$ & $\pm0.26$ & $\pm0.017$ & $\pm0.05$ & $\pm0.04$ & $\pm0.02$ & - & - & - & - \\
   
   

	\hline
\end{tabular}}
\label{tab:photometry}
\end{table*}

\section{Conclusion}

We present here the results of a comprehensive survey of the environment of NGC2683, a Local Volume spiral galaxy using HSC-g band images, extending beyond the Virial radius. A thorough visual search of these images reveals four new satellite galaxy candidates of the NGC2683 system (two of which were independently reported in \cite{Carlsten_2022}), in addition to three confirmed satellite galaxies and five still unconfirmed candidates found in previous surveys. Our deep imaging survey is 100 percent complete out to the $380$\,kpc, well beyond the estimated Virial radius of $220$\,kpc down to an absolute magnitude of $M_{\text{g}}<-11$, and 50 percent complete for $-11<M_{\text{g}}<-9.5$, excluding compact objects of $r_{\text{e}}<300$\,pc.

In our satellite galaxy sample, we note the prevalence of dwarf elliptical or spheroidal type satellite galaxies. This may be a product of our conservative approach in which we have discounted the majority of potential irregular/BCD type galaxies due to the inherent difficulty in discriminating asymmetric dwarf galaxies with background spiral/starburst galaxies. Regardless, this is likely to be the nature of the NGC2683 system and the expected outcome. We expect high density galactic environments, such as satellite galaxy systems within the virial radius of a host galaxy to be mostly populated by dE or dSph galaxies \citep{Houghton_2015, Habas_2020}. As an example, the LMC and SMC of the Milky Way are thought to be coupled and having fallen into orbit around the Milky Way relatively recently rather than forming in-situ \citep{Kallivayalil_van_2013}. Thus if we then ignore the LMC and SMC, we find that nearly the entire satellite system of the Milky Way is dominated by dSph type galaxies. In this scenario, the satellite composition of the NGC2683 system is not atypical. This stands in contrast with results from the the SAGA Survey which suggests that star-forming satellites around Milky Way like hosts are more common than observed in the Local Group \citep{Mao2021}. It is atypical however, that KK69, dw 0846+33 and dw0846p3300 are so diffuse. All three candidates could reasonably be classified as Ultra Diffuse Galaxies, with surface brightnesses much lower than most dwarf galaxies, though dw 0846+33 falls just short of the classification condition. KK69 and dw0846p3300 are also disturbed galaxies, with smooth but irregular morphology.

We also find that the NGC2683 system is moderately sparse; it's total number of detected satellites to an absolute magnitude of $M_g\approx-9.5$ of 12 is less than more populated systems such as M31. This number may drop even further, once candidates are confirmed/unconfirmed, as our sample contains 9 unconfirmed candidates. Given recent efforts to identify satellite candidates of the semi-isolated NGC253 system have revealed a similarly sparse system of satellites \citep{MartnezDelgado2021} and that accretion is understood to primarily build satellite systems, the sparse population of the NGC2683 system given it's isolation is expected. In all, this system compares well to the Milky Way system in terms of the number and morphology (excluding the LMC and SMC) of satellites. Where NGC2683 appears to be somewhat unique, is its absence of any luminous satellite galaxies with magnitude $M_{g}<-15$, resulting in a magnitude gap of $6.1\,m_{g}$ between the host and the brightest satellite. This only occurs in around 5\% of systems in the\emph{TNG100-1} simulation. A more thorough study may in future may be able to draw more insights from this aspect of the NGC2683 environment and investigate the causes leading to large magnitude gaps.

Finally, the 2-dimensional distribution of known satellites and our new sample are consistent with an anisotropic distribution of the same number of satellites, as a <1 percent outlier of the isotropic distribution. This is however reliant on the confirmation of existing candidates some of which are outliers. The 2D axis ratio of the NGC2683 system is $0.26\pm0.16$ and for isotropic systems with the same number of satellite galaxies, it is $0.72\pm0.14$, thus the NGC2683 system is a $\sim2.9\sigma$ outlier. Additionally, the major axis of the ellipse of satellites is approximately aligned with the polar axis of NGC2683. This compares well to MATLAS galaxies where a flattened distribution of satellites is detected \citep{Heesters2021}. We compared the anisotropy of the NGC2683 system to similar (in both host galaxy mass, host galaxy isolation, and number of detected satellites) simulated satellite environments in the IllustrisTNG \textit{TNG100-1} simulation and found the 2D projected axis ratio distribution of these galaxies to be $0.44\pm0.23$, such that the NGC2683 system of satellites is comparable to simulations in this manner. With this limited information it is indicated the NGC2683 system is anisoptric but follow-up observations and research into the 3D and velocity distribution of these candidates is required to robustly compare this system with $\Lambda$CDM simulations on all points of contention, those being co-rotating and highly confined 3D planes, which are not observed in simulations.

\section*{Acknowledgements}
Ethan Crosby, Helmut Jerjen and Markus Dirnberger acknowledge financial support from the Australian Research Council through the Discovery Project DP150100862.
Oliver Müller is grateful to the Swiss National Science Foundation for financial support under the grant number  	PZ00P2-202104. 
Marcel. S. Pawlowski acknowledges funding of a Leibniz-Junior Research Group (project number J94/2020) via the Leibniz Competition, and also thanks the Klaus Tschira Stiftung and German Scholars Organization for support via a KT Boost Fund.
This research has made use of the NASA/IPAC Extragalactic Database, which is funded by the National Aeronautics and Space Administration and operated by the California Institute of Technology. It made also use of the SIMBAD database, operated at CDS, Strasbourg, France.


\section*{Data Availability}
The data underlying this article will be shared on reasonable request to the corresponding author.

Additional data underlying this article were derived from sources in the public domain and are available at: https://ned.ipac.caltech.edu/, https://simbad.u-strasbg.fr/simbad/ and https://hscla.mtk.nao.ac.jp/doc/



\bibliographystyle{mnras}
\bibliography{NGC2683.bib} 


\clearpage 
\appendix

\section{KK69 Photometry} \label{appendix_a}

As described in Section \ref{photometric_modelling} light profiles of dwarf spheroidal/elliptical galaxies are well fitted by a single S\'ersic profile. However, for KK69 specifically with irregular morphology, we firstly fit the galaxy (albeit poorly) with a single S\'ersic model. Components of the galaxy remaining in the residual image of this fit were used to determine the placement of new S\'ersic model fits, this is iterated until there is nothing of significance left in the residuals. \textsc{galfit} is capable of fitting overlapping models with increased computation time. This approach allows us to accurately measure the photometric parameters, in particular the apparent magnitude, even for irregular galaxies with a small margin of error using a process that typically fails to account for asymmetric morphologies.

\begin{figure*}
\includegraphics[draft=false,width=15cm]{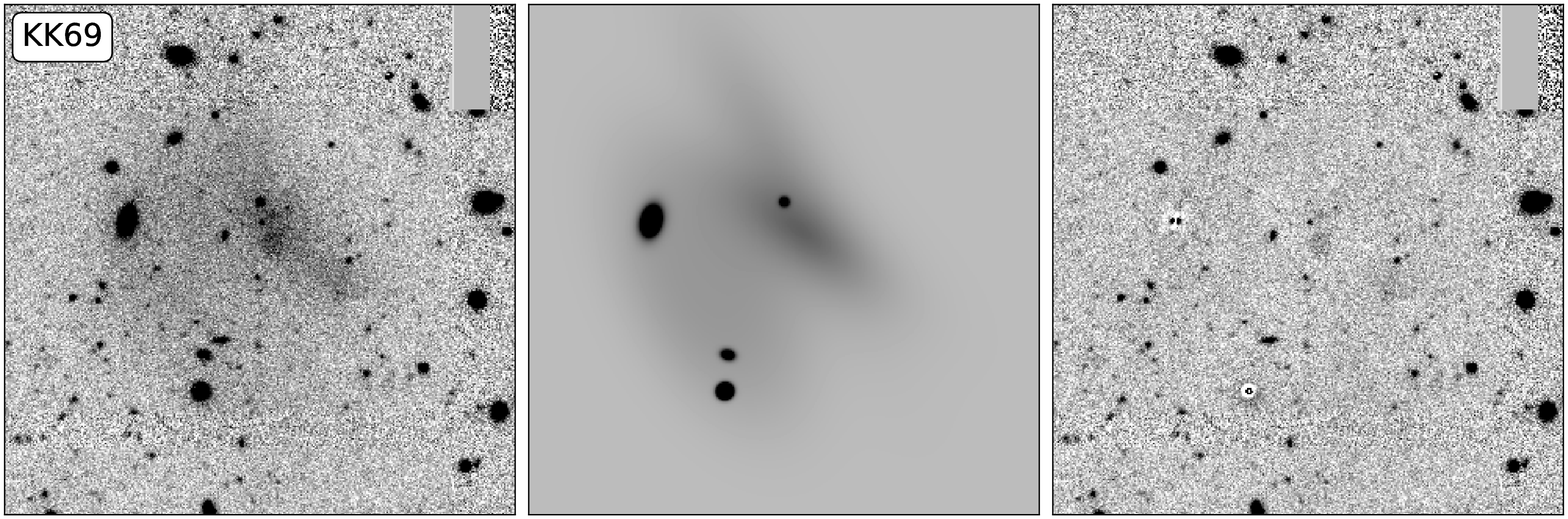}
\caption{\textsc{galfit} results for our KK69. This contains the main co-added HSC-g exposure (left); the \textsc{galfit} model (middle); residual image obtained by subtracting the model from the main exposure (right).}
\label{fig:KK69_phot}
\end{figure*}

While galactic structural parameters in smooth and symmetric galaxies are well defined as the parameters of the single S\'ersic fit, structural parameters are poorly defined in irregular galaxies fit with multiple overlapping models. So for smooth single model galaxies the parameters presented in Table \ref{tab:photometry} are the parameters of the best fit model. While for irregular galaxies these parameters are calculated as follows: the listed S\'ersic index $n$, effective radius $r_e$, and the axis ratio $a/b$ are that of the single S\'ersic profile fit, admitting that said fit may be poor. While the total magnitude is calculated from the combined magnitudes of all the fitted components assumed to be a part of the satellite galaxy from the multi-component model. The final structural parameters for KK69 described here are presented Table \ref{tab:photometry_kk69} and the fitted model shown in Figure \ref{fig:KK69_phot}.

\begin{table*}
\caption{HSC \textit{HSC-g} band photometry and structure parameters of KK69 as in our analysis. Photometric properties are derived from the best-fitting \textsc{galfit} models. Column definitions are described in Table.\ref{tab:photometry}. Magnitudes are in the \textit{HSC-g} band.}
\medskip
\resizebox{\textwidth}{!}{%
\tabcolsep=2pt
\begin{tabular}{*{12}{c@{\hspace{1em}}}c} 
	\hline
    ID & R.A. (J2000) & DEC (J2000) & Type & $m_{g}$ & $A_g$ & $M_g$ & $r_e$ & $r_e$ & $\langle\mu_e\rangle$ & $n$ & $b/a$ \\
    
    - & degrees & degrees & - & mag & mag & mag & arcsec & kpc & mag $\text{arcsec}^{-2}$ & - & - \\
    
    (1) & (2) & (3) & (4) & (5) & (6) & (7) & (8) & (9) & (10) & (11) & (12) \\
    
    \hline
   KK69 & 133.198 & 33.729 & UDG/dT & $15.46 \pm 0.03$ & 0.09 & $-14.50 \pm 0.07$ & $76.7 \pm 0.4$ & $2.07 \pm 0.09$ & $25.66 \pm 0.03$ & $0.34 \pm 0.01$ & $0.83 \pm 0.01$  \\
	\hline
\end{tabular}}
\label{tab:photometry_kk69}
\end{table*}

\bsp	
\label{lastpage}
\end{document}